\documentstyle[aps,eqsecnum]{revtex}
\begin{document}
\title{\hspace{12.0cm} Preprint WSU-NP-96-12 \protect\\
\vskip 1.5cm
Spurious poles of the axial gauge propagators and
dynamics of the interacting fields} 
\author{ A.  Makhlin}
\address{Department of Physics and Astronomy, Wayne State University, 
Detroit, MI 48202}
\date{August 6, 1966}
\maketitle
\begin{abstract}
The origin of the spurious poles of the gauge field
propagators in the temporal axial and the null-plane gauges is discussed.
The conclusion is that these poles do not require any special
prescription.  They are a manifestation of the fact that that the gauge 
field acquires a static configuration.      
\end{abstract}

\section{Introduction}\label{sec:SN1} 

The issue of spurious poles of the propagators in axial and null-plane 
gauges has been controversial for almost two decades. The absence of
ghosts in these physical gauges makes them very attractive despite the
loss of the manifest Lorentz covariance, which latter is usually recovered
in  calculations of observables. Uncertainty in the treatment of spurious
poles is probably  the predominate reason why many  theorists prefer to
use covariant gauges (and enjoy the relativistic invariance during the
intermediate stages of calculations).  Much effort has been spent in order
to find a universal prescription to handle the spurious poles. The
principal value prescription \cite{PV} and the one of  Leibbrandt and
Mandelstam  \cite{ML} were tested in various calculations of the Wilson
loop up to  fourth order perturbation theory.  The total score of gains
and losses looks approximately equal. Solutions to the problem were also
looked for along the lines of path-dependent formulation; its connection
with the problem of the residual symmetry was also understood
\cite{Ivanov}.  It may look surprising, that  the  object of the
controversy, the propagator of the perturbation theory, is so simple, and
that the problem is not specific for  non-abelian gauge fields, but exists
in the same form in  QED as well \cite{SF}.  All of the above studies
attempted  to solve the problem of spurious poles in the general context
of gauge field theory. 

Spurious poles are safe in practical calculations like, for example, in
computing  cross sections. They  either cancel with the  traces in the
numerators of the matrix elements or are unaccessible  for kinematic
reasons. However, the problem does appear in the less--standard
calculations \cite{McLer}, where the choice of the  prescription may
affect the physical results.   

In this paper we discuss the problem of spurious poles keeping in mind an 
environment where the Lorentz and/or translational symmetry is already
broken by the  actual geometry of the physical process, as in, for
example,  deep inelastic electron-proton scattering or in  central
collisions of hadrons or heavy ions. For these processes, one has a
natural choice  of the axis for the gauge condition. Moreover, use of the
covariant gauges is highly undesirable in these cases. Expecting the
creation of the statistical system one immediately is faced with the
problem of the unphysical ghosts distribution.

Instead of an examination of the consequences of different prescriptions 
for quantum field theory calculations, this paper  concentrates on  the
different ways to derive the propagator in the temporal axial gauge
$A^0=0$, and the null-plane gauge $A^+=0$.  Since the object is primitive,
the focus will be  on the classical aspects of the derivation. There is no
difference between the Green functions of the quantum perturbation theory
and the singular  (fundamental) solutions of the system of Maxwell
equations which describe the interaction of the gauge field with the
classical current.  Causality and the influence of initial and boundary
conditions will be taken as the main  guidelines in the derivation of the
Green functions.  

This investigation of the nature of spurious poles is undertaken as a part
of a general study of  field dynamics in deeply inelastic high  energy
processes \cite{QFK,QGD,WD1,WDG}.   This analysis is conceived as a
prototype  for a calculation of the gluon Green functions in the ``wedge
dynamic''\cite{WD1,WDG}. The final conclusion of this work is that
contributions of spurious poles correspond to  the classical character of
the gauge field. These poles do not need special treatment, but they serve
as important indicators of the  static configuration of the gauge field. 

\section{Gauge field correlators in the gauge $A^{0}=0$}\label{sec:SN2}   

In this section we briefly illustrate our approach for obtaining the 
fundamental solutions of Maxwell equations using a technically more simple
example of the temporal axial gauge $A^{0}=0$. This gauge has the known
problem of the infinite growth of the propagator which manifests itself as
an additional pole of the polarization sum and uncertainty of the
treatment of this pole . However,  because of the manifest 
translational invariance
(the gauge condition singles out a direction and not the point!), one may
use a simple Fourier analysis and avoid specific problems due to
nonlocality of the propagator in the gauge $A^{\tau}=0$.  We shall not
attempt to invert the matrix differential  operator, either algebraically
or by means of the path integral formalism. Instead, we shall obtain
various propagators  of the perturbation theory via their expansion over a
set of classical solutions. Analysis of these solutions sheds light upon
the
meaning of the different prescriptions to handle the gauge poles.          

The gauge invariant action of the Abelian theory  is as follows,
\begin{eqnarray}
{\cal S}=\int {\cal L}(x)d^4x =\int [-{1\over 4}
F^{\mu\nu}(x)F_{\mu\nu}(x) -j^\mu A_\mu ]\sqrt{-{\rm g}} d^4 x~.
\label{eq:E2.1}\end{eqnarray}              
 
Its variation with respect to the gauge field yields the Lagrangian
equations of motion,
\begin{eqnarray}
\partial_\mu F^{\mu\nu}=j^\nu~.
\label{eq:E2.2}\end{eqnarray}    
With the gauge condition $A^{0}=0$, the system of
Maxwell equations with the classical current $j^\mu$ can be written down
in terms of the 3-dimensional Fourier components as follows:
\begin{eqnarray} 
[(\partial_{t}^{2}+ {\bf k}^2)\delta^{ij} - k^i k_j]A^j({\bf k},t)
=j^i({\bf k},t)
\label{eq:E2.3}\end{eqnarray}  
\begin{eqnarray}
{\cal C}(x) = \partial_x E_x+\partial_y E_y+\partial_z E_z-j^0=0~~, 
\label{eq:E2.4}\end{eqnarray}                                          
where the indices $i$ and $j$ numerate the spacial coordinates     
and  $E_i = {\dot A}_i$ is the strength of the electric field.    
Equation (\ref{eq:E2.4}) is a  constraint corresponding to 
Gauss' law.

As mentioned above, the set of the ordered field correlators  
used in the quantum theory ,
\begin{eqnarray}
D^{\mu\nu}_{00}(x,y)=-i\langle 0|T(A^\mu(x)A^\nu(y))|0 \rangle,\;\;\; 
D^{\mu\nu}_{10}(x,y)=-i\langle 0|A^\mu(x)A^\nu(y)|0 \rangle  \nonumber \\
D^{\mu\nu}_{01}(x,y)=-i\langle 0|A^\nu(y)A^\mu(x)|0 \rangle,\;\;\; 
D^{\mu\nu}_{11}(x,y)=-i\langle 0|T^{\dag}(A^\mu(x)A^\nu(y))|0 \rangle ~~, 
\label{eq:E2.5}\end{eqnarray}               
along with the retarded and advanced Green functions
\begin{eqnarray}
D^{\mu\nu}_{ret}(x,y)=D^{\mu\nu}_{00}(x,y)-D^{\mu\nu}_{01}(x,y)=
-i\theta(x^0-y^0)\langle 0|[A^\mu(x),A^\nu(y)]|0 \rangle ~~,   \nonumber \\    
D^{\mu\nu}_{adv}(x,y)=D^{\mu\nu}_{00}(x,y)-D^{\mu\nu}_{10}(x,y)=
i\theta(y^0-x^0)\langle 0|[A^\mu(x),A^\nu(y)]|0 \rangle      
\label{eq:E2.6}\end{eqnarray}            
coincide with various singular or fundamental solutions of the classical
equations and can be studied regardless their quantum nature. The quantum
content reveals itself only when the Fock space is constructed and the
bilinear expansion of these correlators over the eigenfunctions  is
obtained as a quantum average over the vacuum state.  It is also important
that except for the simple linear relations between various correlators 
we have also dispersion relations which reflect causal properties of the
theory,
 \begin{eqnarray}
{\rm Re} D^{\mu\nu}_{ret,adv}(k^0,{\bf k})= {1\over \pi i}
{\cal P}\!\!\!\int_{-\infty}^{\infty} { {\rm Im} 
D^{\mu\nu}_{ret,adv}(\omega,{\bf k}) \over \omega -k^0 } d~\omega~.
\label{eq:E2.7}\end{eqnarray}            
They can be formally derived from (\ref{eq:E2.6}) and the relation,
$D_0=D_{ret}-D_{adv}=2i{\rm Im} D_{ret}$.

\subsection{Wightman functions of the free gauge field} 
\label{subsec:SBS1}   
                                      
Our first goal is to find solutions of the homogeneous system
(\ref{eq:E2.3})  of Maxwell equations (with $j_i({\bf k},t)=0$.) The
solutions of the homogeneous equations will be looked for in terms of the
auxiliary functions,       
\begin{eqnarray}   
\Phi= \partial_{x}A_x +\partial_{y} A_y ,\;\;\; 
\Psi= \partial_{y}A_x -\partial_{x} A_y ,\;\;\;  {\rm and} \;\; 
{\sf A}=A_z~,
\label{eq:E2.8}\end{eqnarray}       
for which one obtains a system of equations,
\begin{eqnarray}   
(\partial_{t}^{2} + {\bf k}_{\bot}^2 +k_{z}^{2})\Psi({\bf k},t)=0~~,
\label{eq:E2.9}\end{eqnarray} 
\begin{eqnarray}  
(\partial_{t}^{2} + k_{z}^{2})\Phi({\bf k},t)
-{\bf k}_{\bot}^2 k_z {\sf A}({\bf k},t) = 0~~, 
\label{eq:E2.10}\end{eqnarray} 
\begin{eqnarray}  
(\partial_{t}^{2} + {\bf k}_{\bot}^{2}){\sf A}({\bf k},t)
- k_{z}^{2} \Phi({\bf k},t) = 0~.
\label{eq:E2.11}\end{eqnarray}           
This nonsymmetric form is consciously chosen in order to mimic
the physical asymmetry of the gauge $A^{\tau}=0$.
Eq.~(\ref{eq:E2.9}) has solutions $e^{-i|{\bf k}|t}$ and $e^{i|{\bf k}|t}$.
The system of equations (\ref{eq:E2.10}) and (\ref{eq:E2.11}) 
has the first integral,
\begin{eqnarray}   
\partial_t (\Phi({\bf k},t)+ k_z {\sf A}({\bf k},t))
\equiv -i {\dot \rho}({\bf k})=0~,
\label{eq:E2.12}\end{eqnarray}            
which expresses the conservation of the constraint with time but does
not coincide with  Gauss' law.  To solve this system it is necessary to
differentiate the equations  once more, thus converting them into the wave
equations for the electric field strength. The third order of these
independent equations agrees with the number of non-vanishing components of 
the vector potential. 
Using of  Eq.~(\ref{eq:E2.12}) immediately leads to the two
equations of third order for the functions $\Phi$ and ${\sf A}$:
\begin{eqnarray}   
(\partial_{t}^{3} + {\bf k}^2\partial_{t})\Phi({\bf k},t)=
-i k_{\bot}^{2} \rho({\bf k})~~, \;\;\;
(\partial_{t}^{3} + {\bf k}^2\partial_{t}) {\sf A}({\bf k},t)=
-i k_{z} \rho({\bf k})~~,
\label{eq:E2.13}\end{eqnarray}            
which have as set of solutions,
\begin{eqnarray}   
\int^t dt'e^{-i|{\bf k}|t'}~,\;\; \int^t dt'e^{i|{\bf k}|t'}~,\;\;\;
 {\rm and}\;\; \rho({\bf k})~t -\rho_1({\bf k})~t_0~~. \nonumber
\end{eqnarray}
The $t$-independent term in the third solution is a pure gauge and
can be omitted. 
After a short exercise in algebra and normalization according to
\begin{eqnarray}   
\int^t d^3 {\bf r} V^{(\lambda) \ast}_{{\bf k},i}({\bf r},t)
i{\stackrel{\leftrightarrow}{\partial}}_{t} 
V^{(\lambda)}_{{\bf k},i}({\bf r},t) =\delta_{\lambda\lambda'}~,
\delta({\bf k}-{\bf k'})
\label{eq:E2.14}\end{eqnarray} 
one obtains three modes: two orthogonal radiation modes,
\begin{eqnarray}  
V^{(1)}_{{\bf k}}(x)={1 \over (2\pi)^{3/2}(2|{\bf k}|)^{1/2}}
\left( \begin{array}{c} 
                         k_y/|{\bf k}| \\ 
                        -k_x/ |{\bf k}| \\ 
                         0 
                             \end{array} \right)
e^{-ikx}~~,      
V^{(2)}_{{\bf k}}(x)={1 \over (2\pi)^{3/2}(2|{\bf k}|)^{1/2}}
\left( \begin{array}{c} 
                         k_x k_z /|{\bf k}_{\bot}|^{2}  \\ 
                         k_y k_z /|{\bf k}_{\bot}|^{2} \\ 
                         -1
                             \end{array} \right) e^{-ikx}~~,      
\label{eq:E2.15} 
\end{eqnarray} 
which obey Gauss law without the charge, and a Coulomb mode,
\begin{eqnarray} 
V^{(3)}_{{\bf k},i}(x)={ i k_i \rho({\bf k}) \over {\bf k}^2}e^{i{\bf kr}}~t ~.
\label{eq:E2.16}\end{eqnarray}    
The norm of the Coulomb mode, (as defined by Eq.~(\ref{eq:E2.14})), equals
zero, and this mode is orthogonal to $V^{(1)}$ and $V^{(2)}$. One can
easily write down the coordinate form of this solution,
\begin{eqnarray}  
V^{(3)}({\bf r},t)= - t {\partial \over \partial x^i}
\int {\rho({\bf r'})  \over |{{\bf r}-{\bf r'}| }} d {\bf r'}~,
\label{eq:E2.17}\end{eqnarray}     
which is negative of the Coulomb field strength times $t$. Though this
solution  obeys the equations of motion without the current, it does not
obey Gauss' law without a charge. Therefore, it should be discarded in the
decomposition of the radiation field. However, it should have been kept
when or if the radiation field in the presence of the static source
$j^0({\bf k})=  \rho({\bf k})$ is  considered.                              

With the two radiation modes in hand, we can obtain the field correlators
of the radiation field. The sum over the two physical polarizations invokes
a transverse projector
\begin{eqnarray}  
d^{ij}(k)=\sum_{\lambda} \varepsilon^{i}_{(\lambda)}(k) 
 \varepsilon^{j}_{(\lambda)}(k) =\delta^{ij}-{k^i k^j \over {\bf k}^2}~~.
\label{eq:E2.18}\end{eqnarray}       
We thus see that an artificial asymmetry of the structure of the modes  is
washed out. The final non-covariant expressions for the Wightman functions
are as follows:
\begin{eqnarray}
D^{ij}_{10}(x,y)=-2\pi i\int {d^3 {\bf k} \over 2 |{\bf k}|(2\pi)^3}
d^{ij}(k) e^{-i|{\bf k}|t+i{\bf kr}}, \;\;\;  
D^{ij}_{01}(x,y)=-2\pi i\int {d^3 {\bf k} \over 2 |{\bf k}|(2\pi)^3}
d^{ij}(k) e^{+i|{\bf k}|t+i{\bf kr}}~,  
\label{eq:E2.19}\end{eqnarray}           
and we shall delay their rewriting in the formal covariant form until
the propagators are found.      

\subsection{Green functions of the gauge field} 
\label{subsec:SBS2}   

We now consider the interaction of the gauge field with the classical source.
The first way to obtain the Green function is more or less formal and
not general since it relies on the Fourier analysis in terms of plane 
waves.
Projecting the field and the current vectors in the system (\ref{eq:E2.3}) 
onto the two ``orthogonal'' directions, {\em i.e.} 
\begin{eqnarray}  
     A^{(tr)}_{i}=(\delta_{ij} - k_i k_j/ {\bf k}^2)A_j~,\;\;
     A^{(L)}_{i}=(k_i k_j/ {\bf k}^2)A_j~ ,
\label{eq:E2.20}\end{eqnarray}              
we arrive at two different ordinary differential equations for the
radiation and the longitudinal fields:
\begin{eqnarray} 
[(\partial_{t}^{2}+ {\bf k}^2)A^{(tr)}_{i}({\bf k},t)
=j^{(tr)}_{i}({\bf k},t)~~,
\label{eq:E2.21}\end{eqnarray} 
and   
\begin{eqnarray} 
\partial_{t}^{2} A^{(L)}_{i}({\bf k},t)
= j^{(L)}_{i}({\bf k},t)~~.
\label{eq:E2.22}\end{eqnarray}    
The fundamental solutions of these equations corresponding to the retarded
solutions are known, 
\begin{eqnarray}  
     A^{(tr)}_{i}({\bf k},t)= \int_{-\infty}^{+\infty} \theta (t-t')
{\sin |{\bf k}|(t-t') \over |{\bf k}|} j^{(tr)}_{i}({\bf k},t') dt'~ ,
\label{eq:E2.23}\end{eqnarray} 
\begin{eqnarray}  
     A^{(L)}_{i}({\bf k},t)= \int_{-\infty}^{+\infty} \theta (t-t')
(t-t') j^{(L)}_{i}({\bf k},t') dt'~ ,
\label{eq:E2.24}\end{eqnarray}              
After the Fourier transformation over  time, {\em i.e.}  in the full 
energy-momentum representation,we find that
\begin{eqnarray}  
     A^{(tr)}_{i}({\bf k},\omega)= {1 \over (\omega +i0)^2 - {\bf k}^2}
    (\delta^{il}-{k^i k^l \over {\bf k}^2} ) j^l({\bf k},\omega)~,
\label{eq:E2.25}\end{eqnarray} 
\begin{eqnarray}  
A^{(L)}_{i}({\bf k},\omega)= {1 \over (\omega +i0)^2 }
    {k^i k^l \over {\bf k}^2} j^l({\bf k},\omega)~.
\label{eq:E2.26}\end{eqnarray}             
We have thus obtained the 
$(\omega+i0)$--prescription for the poles               
for both  transverse and longitudinal modes  which  guarantees the retarded
character of the response.  Correspondingly, for the poles of the advanced 
propagator we now must obtain the $(\omega-i0)$--prescription. 
Conbining Eqs.~(\ref{eq:E2.25}) and (\ref{eq:E2.26}) together we find a 
familiar form of the axial gauge propagator,
\begin{eqnarray}
D_{ret}^{ij}({\bf k},\omega) = 
{d^{ij}({\bf k}) \over (\omega +i0)^2 - {\bf k}^2}
+{k^i k^j \over (\omega +i0)^2 {\bf k}^2}
={1 \over (\omega +i0)^2 - {\bf k}^2} (\delta^{ij} 
-{k^i k^j \over (\omega +i0)^2})~.                   
\label{eq:E2.27}\end{eqnarray}           
The full covariant form is obtainable according to the following rule
of replacements: 
\begin{eqnarray}  
k^i \rightarrow k^\mu - u^\mu (ku),~~~ 
\delta^{ij} \rightarrow -g^{\mu\nu} + u^\mu u^\nu ,~~~
\omega\rightarrow ku,~~~ 
{\bf k}^2 \rightarrow  (ku)^2 -k^2 ~~.  \nonumber
\end{eqnarray}        
Then we immediately get:    
\begin{eqnarray} 
D_{ret}^{\mu\nu}(k) = 
{d^{\mu\nu}({\bf k}) \over (\omega +i0)^2 - {\bf k}^2}, \nonumber \\
 d^{\mu\nu }(k)=-g^{\mu\nu }+{k^\mu u^\nu + u^\mu k^\nu  \over ku }
- {k^\mu k^\nu   \over (ku)^2 }~~.
\label{eq:E2.28}\end{eqnarray}       
                                   
Considering this fully covariant form of the propagator as a given, we
immediately run into several problems which manifest themselves through 
the poles of the projector $d^{\mu\nu }(k)$.
 
(1) The projector $d^{\mu\nu }(k)$ has no first order poles at $ku=0$.
Unlike the two first order poles at $\omega =\pm |\bf k|$ in the
transverse modes, we have one second order pole in the longitudinal mode.
The residue in this pole is therefore given by the derivative of the
integrand. Consequently, one obtains the term with the linear time
dependence of the longitudinal constituent of the vector potential, 
corresponding to the static configuration of the electric field, which is
not a subject of quantum dynamics.

(2) Disposition of the poles of the transverse modes at  $\omega =\pm
|{\bf k}|-i0$  eventually leads to the propagator which explicitly
exhibitss the proper light-cone behavior, including the Lorentz invariant
definitions of ``before'' and ``after.'' The absence of the $|\bf k|^2$
(Laplasian) in  Eq.~(\ref{eq:E2.22}) means that the longitudinal field is
not propagating. Consequently, the  $(\omega\pm i0)$-prescriptions are not
Lorentz invariant for these modes and even for the retarded and advanced
propagators, they are misleading. 

(3) Consider now the difference $2D_{0}=D_{ret}-D_{adv}$, which obeys the
homogeneous equation and, therefore, cannot contain the longitudinal modes
in its decomposition. However, 
\begin{eqnarray}  
{1\over (\omega+i0)^2} -{1\over (\omega-i0)^2} = 
2\pi i \delta' (\omega)\neq 0 ~.\nonumber
\end{eqnarray}        
Therefore, the commutator of the free field acquires an unphysical 
contribution from the longitudinal field which is a  remnant  of the
improper handling the gauge poles. From this point of view and in order to
get agreement with the dispersion relations, the principal value
prescription looks the most attractive, because the Coulomb part of the
propagator will not contribute to its imaginary part. However, it is not a
physical solution of the problem.

(4) Since projections of Eq.~(\ref{eq:E2.3}) onto  Eqs.~(\ref{eq:E2.21})
and ~(\ref{eq:E2.22}) are orthogonal, we may  use  current conservation
and rewrite  Eq.~(\ref{eq:E2.22}) as
\begin{eqnarray}  
\partial_t {\rm div} {\bf E}+ {\rm div}{\bf j} = 
\partial_t ({\rm div} {\bf E}-j^0)=0~.   
\label{eq:E2.29}\end{eqnarray}  
This is now in the form of the equation of the constraint conservation. 
Therefore, there
is no  reason to integrate it in the ``retarded'' or  ``advanced''
manner. Moreover, there is no physically motivated prescription for the
integration which recovers the potential ${\bf A}$ via the electric field
${\bf E}={\bf{\dot A}}$.

All these problems  appear if we obtain the propagator formally inverting
the differential operator of the initial system Eq.~(\ref{eq:E2.3}).
These problems cannot be resolved until the origin of every pole is traced.  

\subsection{Straightforward integration of the field equations} 
\label{subsec:SBS3}   

Solution of the Cauchy problem for the free radiation field meets with no
difficultiess. Therefore we must find a way of obtaining the solution of the
inhomogeneous equations which would properly treat the static solutions
as a certain limit of the full emission problem. For this purpose, we shall
rewrite the system of Maxwell equations in the following form,
\begin{eqnarray}   
(\partial_{t}^{2} + {\bf k}_{\bot}^2 +k_{z}^{2})\Psi({\bf k},t)=
i[k_xj_y({\bf k},t)-k_yj_x({\bf k},t)]\equiv j_\psi ({\bf k},t)~~,
\label{eq:E2.30}\end{eqnarray} 
\begin{eqnarray} 
(\partial_{t}^{2} + k_{z}^{2})\Phi({\bf k},t)
-i{\bf k}_{\bot}^2 k_z {\sf A}({\bf k},t) = 
i[k_xj_x({\bf k},t)+k_yj_y({\bf k},t)]\equiv j_\phi ({\bf k},t)~~, 
\label{eq:E2.31}\end{eqnarray} 
\begin{eqnarray} 
(\partial_{t}^{2} + {\bf k}_{\bot}^{2}){\sf A}({\bf k},t)
+ i k_z \Phi({\bf k},t) = j_z({\bf k},t)~,
\label{eq:E2.32}\end{eqnarray}                  
and attempt to find a partial solution by means of the ``variation of 
parameters'' method. As in the case of the homogeneous system in section
\ref{subsec:SBS1}, it is expedient to differentiate Eqs.~(\ref{eq:E2.31}) 
and ~(\ref{eq:E2.32}) once more. The difference of the resulting equations
gives an equation of the constraint conservation. The equation of
constraint can be integrated as follows,
\begin{eqnarray}   
{\dot\Phi}({\bf k},t)+ ik_z {\sf{\dot A}}({\bf k},t))
= [j^0({\bf k},t) - \rho({\bf k})]~~,
\label{eq:E2.33}\end{eqnarray}       
where $\rho({\bf k})$  is  a constant of integration and, until it is
set equal to zero, the equation of Gauss' law is not explicitly used.
Eq.~(\ref{eq:E2.33}) allows one to obtain two independent equations instead
the system  (\ref{eq:E2.31})--(\ref{eq:E2.32}):
\begin{eqnarray}   
(\partial_{t}^{2} +{\bf k}^{2}){\dot\Phi}({\bf k},t)
=  k_{\bot}^{2}[j^0({\bf k},t) - \rho({\bf k})]+\partial_t 
j_\phi ({\bf k},t)\equiv f_\phi ({\bf k},t)~~, 
\label{eq:E2.34}\end{eqnarray} 
\begin{eqnarray}  
(\partial_{t}^{2} + {\bf k}^{2}){\dot{\sf A}}({\bf k},t)
= -i k_{z}[j^0({\bf k},t) - \rho({\bf k})]-\partial_t 
j_z({\bf k},t) \equiv f_z({\bf k},t)~.
\label{eq:E2.35}\end{eqnarray}                  
Varying the constants in the decomposition of the partial solution
we find that
\begin{eqnarray} 
{\cal F}={i \over 2|{\bf k}|} 
\bigg( e^{-i|{\bf k}|t}\int_{-\infty}^{t} d~t'f(t')e^{i|{\bf k}|t'} 
- e^{i|{\bf k}|t}\int_{-\infty}^{t} d~t'f(t')e^{-i|{\bf k}|t'} \bigg) ~~,
\label{eq:E2.36}\end{eqnarray}                            
where ${\cal F}$ stands for $\Psi$, ${\dot\Phi}$ or ${\dot{\sf A}}$
and $f$ stands for the R.H.S. of either (\ref{eq:E2.34})  or
(\ref{eq:E2.32}).  

Starting from this point, one may wish to take a short cut and find 
the solution by means of the symbolic Fourier calculus. Using the
Fourier representation for the source, $f$,~ and the time-independent
integration constant $\rho$,
\begin{eqnarray} 
f({\bf k},t)=\int_{-\infty}^{\infty} {d\omega \over 2\pi}
f({\bf k},\omega)e^{-i\omega t}, \;\;\;
\rho({\bf k})=\int_{-\infty}^{\infty} d\omega 
\rho({\bf k}) \delta(\omega)e^{-i\omega t}~~,   
\label{eq:E2.37}\end{eqnarray}                            
one obtains for the magnetic field $\Psi$ of the transverse mode
\begin{eqnarray} 
\Psi ({\bf k},t)=\int_{-\infty}^{\infty} {d\omega \over 2\pi}
{j_\psi ({\bf k},\omega) \over (\omega +i0)^2 -{\bf k}^2} e^{-i\omega t}~~,   
\label{eq:E2.38}\end{eqnarray}       
and for the components of the electric field,
\begin{eqnarray} 
{\dot \Phi}({\bf k},t)=\int_{-\infty}^{\infty} {d\omega \over 2\pi}
{k_{\bot}^{2}[j^0({\bf k},\omega) - 2\pi\delta (\omega) \rho({\bf k})] 
-i\omega j_\phi ({\bf k},\omega) 
\over (\omega +i0)^2 -{\bf k}^2}   e^{-i\omega t}~~,   
\label{eq:E2.39}\end{eqnarray}  
and  
\begin{eqnarray} 
{\dot {\sf A}}({\bf k},t)=\int_{-\infty}^{\infty} {d\omega \over 2\pi}
{-i k_z[j^0({\bf k},\omega) -2\pi\delta (\omega) \rho({\bf k})] 
-i\omega j_z ({\bf k},\omega) 
\over (\omega +i0)^2 -{\bf k}^2}   e^{-i\omega t}~~.   
\label{eq:E2.40}\end{eqnarray}             
The integration which has led to Eqs.~(\ref{eq:E2.38})-(\ref{eq:E2.40})
gives the electric and magnetic fields and thus is retarded. These
equations do not have  poles at $\omega =0$. However,
we still have to integrate these equations once more in order to
find the vector potential of the gauge field. This  results in
an $\omega$ appearingin the denominator and yields the final answer,
\begin{eqnarray} 
A_i({\bf k},t)=\int_{-\infty}^{\infty} {d\omega \over 2\pi}
{e^{-i\omega t} \over (\omega +i0)^2 -{\bf k}^2} [
 k_i{j^0({\bf k},\omega) -2\pi\delta (\omega) \rho({\bf k})\over \omega} 
  + j_i ({\bf k},\omega)] ~~,   
\label{eq:E2.41}\end{eqnarray}  
again, without any motivated prescription for the new pole.  In order to
obtain the already known expression for the propagator we must set  the
constant of integration $\rho({\bf k})$ equal to zero, thus explicitly
incorporating Gauss' law. Next, it is expedient to use  current
conservation and to rewrite the Fourier component of the charge density
as $j^0({\bf k},\omega)=-k^ij_i({\bf k},\omega)/\omega$. In this way, we
immediately obtain Eqs.~(\ref{eq:E2.27})-(\ref{eq:E2.28}), but without 
any physical handle on the second order pole at $\omega =0$.  However, if 
$j^0({\bf k},\omega) \sim \delta (\omega)$ then $k_ij^i({\bf
k},\omega)=0$, and a formal usage of  current conservation in the
Fourier representation becomes ambiguous. Formally, the problem manifests
itself through the ambiguity of the function $\omega^{-1}\delta (\omega)$.
In this case, one should return to the 
Eqs.~(\ref{eq:E2.39})-(\ref{eq:E2.40}), perform the $\omega$-integration
using the $\delta(\omega)$, and end up with the linear dependence of the
vector potential on time $t$. 
The loss  of continuity in the description of the limit static case which
shows up here is very unlikely. 
Unless we are dealing with the canonical
scattering problem, the physics associated with the static fields is
indeed important. 
The amount of mathematical ambiguities that have appeared in the last few
lines is more than sufficient to show that it is better not to take the
short cut by using the Fourier picture. Let us proceed more gradually
and continue the integration of the time variables. 

Eq.~(\ref{eq:E2.36}) with ${\cal F}=\Psi$ and $f=j_\psi$ already gives
the solution in quadrature  form. To obtain $\Phi$ and ${\sf A}$,
one should integrate twice, {\em e.g.},
\begin{eqnarray} 
{\sf A}({\bf k},t_1)={i \over 2|{\bf k}|} 
\bigg( \int_{-\infty}^{t_1} d~t'~e^{-i|{\bf k}|t'}
\int_{-\infty}^{t'} d~t_2 ~e^{i|{\bf k}|t_2}~f_z(t_2)
 -\int_{-\infty}^{t_1} d~t' ~e^{i|{\bf k}|t'}
\int_{-\infty}^{t'} d~t_2 ~e^{-i|{\bf k}|t_2}~f_z(t_2) \bigg) ~.
\label{eq:E2.42 }\end{eqnarray}   
Of the two integrations here, the first one, $dt_2$, recovers the electric
field, ${\dot{\sf A}}$,  via the source $f_z$, while the second integration,
$dt'$, is used to find the potential ${\sf A}$. It is easy to see 
that this integration is held to the limits $ t_2 < t' < t_1 $ and thus can 
be done first;  
\begin{eqnarray} 
{ \sf A } ({\bf k},t_1)= 
{ -1 \over |{\bf k}| } 
 \int_{-\infty}^{t_1} d~t_2
\bigg({\cos |{\bf k}|(t_1-t_2) \over |{\bf k}| } -{1 \over |{\bf k}| } \bigg) 
[ i k_{z} j^0({\bf k},t_2) -\partial_t j_z({\bf k},t_2)] ~~,
\label{eq:E2.43}\end{eqnarray}         
where the constant $\rho({\bf k})$ of the constraint integration is
temporarily burried into $j^0({\bf k},t)$.  In fact, the function
$\theta (t_1-t_2) |{\bf k}|^{-1}[\cos |{\bf k}|(t_1-t_2) -1]$ is exactly the
retarded Green function of the ordinary differential equations 
(\ref{eq:E2.34}) and (\ref{eq:E2.35}). Thus, it could be used immediately
to obtain the solution. In this way, one cannot trace the origin of the
spurious pole.  

Every term on the RHS of Eq.~(\ref{eq:E2.43}) should be integrated by
parts with the assumption that the sources vanish as $t\rightarrow
-\infty$.  After the time derivatives of $j^0$ are replaced by the
divergence of the current,  $\partial_t j^0({\bf k},t) =-i k_l j^l({\bf
k},t)$, and the same calculations are repeated for the function $\Phi$, we
arrive at the final answer, 
\begin{eqnarray}  
     A_{i}({\bf k},t_1)= \int_{-\infty}^{t_1} 
{\sin |{\bf k}|(t_1-t_2) \over |{\bf k}| } 
\bigg( \delta_{il}-{k_i k_l\over {\bf k}^2 }\bigg)j^{l}({\bf k},t_2) dt_2
-\int_{-\infty}^{t_1} t_2 {k_i k_l\over {\bf k}^2 } j^l({\bf k},t_2) dt_2
-{k_i\over i{\bf k}^2 } t_1 j^0({\bf k},t_1)~ ,
\label{eq:E2.44}\end{eqnarray} 
Recalling the expression (\ref{eq:E2.16}) for the field of a static charge, we
see that the last term in Eq.~(\ref{eq:E2.44}) represents an 
instantaneous  distribution of the potential at the moment $t_1$,
corresponding to the charge density taken at that same moment. Remembering
that the charge density $j^0({\bf k},t_1) $ in Eq.~(\ref{eq:E2.44}) still
includes the arbitrary constant $\rho({\bf k})$, we see that  imposing the
constraint indeed affects only the potential of static charge distribution
and puts it in agreement with Gauss' law.  Although the proper status
of the static field in the Green function has been recovered, the last
two terms of Eq.~(\ref{eq:E2.44}) have
lost an explicit translational invariance. To restore it, one
should rewrite the last term as 
\begin{eqnarray}  
{k_i\over i{\bf k}^2 }t_1  \int_{-\infty}^{t_1} dt_2
{d j^0({\bf k},t_2) \over dt_2}~=
{ k_i k_l \over  {\bf k}^2 } t_1  \int_{-\infty}^{t_1} 
j^l({\bf k},t_2)~ , \nonumber
\end{eqnarray} 
which is meaningful only if the source is not entirely static. After that,
one obtains propagator in the form given earlier by the 
Eqs.~(\ref{eq:E2.23}) and (\ref{eq:E2.24}). The subsequent Fourier
transformation leads to the second order pole at $\omega =0$ in the
longitudinal part of the propagator.
                                         
Two points from the above discussion are the most essential. First, the
desired prescription is not found even for the retarded and  advanced
propagators, even in the case when we may rely on the most powerful arguments
coming from analyticity and causality. Second, the Wightman functions
(solutions of the homogeneous Maxwell equations) are built entirely from
the free radiation fields and do not have these poles at all. Thus,
the spurious poles cannot acquire any prescription in the $T$--ordered
(Feynman) Green functions  either.  Actually, these poles are the price
we pay for the loss of control over the dynamics of the longitudinal
field when it approaches  the static limit.

\section{Propagator of the null-plane gauge}\label{sec:SN3}  

The null-plane gauge $A^+=A^0+A^3=0$ is a constituent part of the
null-plane dynamics which uses the light-like direction $x^+=x^0+x^3$ as
the  direction of the dynamical evolution. 
Physically, one should think of this gauge as  the
limit of the temporal axial gauge  $uA=0$, when the  4--vector
$u^\mu$ is of the form: $u^\mu = (\cosh y, {\bf 0},\sinh y)$. The vector 
$u^\mu$ can be thought of as the velocity of the proton  and  it is normal
to the space-like hyperplane where the observables identifying the proton
are defined.  Geometrically, this plane is almost parallel to the
null-plane in the limit of $y\rightarrow\pm\infty$. In this limit, the
normal and tangential directions become almost degenerate; however, we shall
keep this difference in mind. If $y\rightarrow +\infty$, then  the Lorentz
contracted proton is confined in the $xy$ plane which moves with rapidity
$y$ in the positive $z$-direction. The gauge condition becomes 
$uA\approx e^y(A^0-A^3)/2=0$, $A^- \approx 0$.  If $y\rightarrow
-\infty$, then the proton is moving in the negative  $z$-direction and the
gauge condition changes to  $uA\approx e^{-y}(A^0+A^3)/2=0$, $A^+
\approx 0$.    Thus we obtain two null-plane gauges as the limit of the
temporal axial gauge; however, in a correspondence which is opposite to
what is naively expected.  This important fact finds a natural physical  
explanation in the scope of the wedge form of Hamiltonian dynamics
\cite{WD1,WDG}. Some auxiliary arguments are submitted at the end
of this section. For the present
discussion, it is important that the spurious  pole of the polarization
sum of the null-plane gauges, 
\begin{eqnarray} 
 d^{\mu\nu }(k)=-g^{\mu\nu }+{k^\mu n^\nu + n^\mu k^\nu  \over kn }~~,
\label{eq:E3.1}\end{eqnarray}
as it was in the case of temporal axial gauge,       
originates from the longitudinal constituent of the gauge field. 
Here, $n^\mu$ has only one non-zero component, either $n^-$ or $n^+$.

Mathematically, the  light-like limit of the time-like direction is 
always singular. Thus, for the sake of safety, it is expedient to start
from the very beginning. The metric tensor has the following nonvanishing
components, $g_{+-}=g_{-+}=1/2$,  ~$g^{+-}=g^{-+}=2$,  and
~$g_{rs}=g^{rs}=-\delta_{rs}$.  Here, ~$r,s=1,2$~ label  the $x$- and
$y$-coordinates. Tree components of the potential, $A_+=A^-/2$, $A_r$, 
are canonical coordinates, and have the electric fields,
$E^-=-F^{+-}/2=-2\partial_-A_+$  and $E^r =-F^{+r}/2=2\partial_-A_r$, as
the canonical momenta.  Maxwell equations can be conveniently
rewritten in terms of three functions, 

\begin{eqnarray}   
\Phi= \partial_{x}A_x +\partial_{y} A_y ,\;\;\; 
\Psi= \partial_{y}A_x -\partial_{x} A_y ,\;\;\;  {\rm and} \;\; 
A_+ \equiv 2A^-.
\label{eq:E3.2}\end{eqnarray} 
with the  sources,      
\begin{eqnarray}   
j^\phi= \partial_{x}j^x +\partial_{y} j^y ,\;\;\; 
j^\psi= \partial_{y}j^x -\partial_{x} j^y ,\;\;\; j^-\equiv 2j_+~~ 
{\rm and} \;\; j^+ \equiv 2j_-
\label{eq:E3.3}\end{eqnarray}       
on the right hand side:
\begin{eqnarray}   
(4\partial_{+}\partial_{-} -\nabla_{\bot}^{2}) \Psi=-j^\psi~, 
\label{eq:E3.4}\end{eqnarray}  
\begin{eqnarray}   
4\partial_{+}\partial_{-}\Phi -2\nabla_{\bot}^{2}\partial_{-}A_+
=-j^\phi~,
\label{eq:E3.5}\end{eqnarray}
\begin{eqnarray}   
4\partial_{+}\partial_{-}A_+ -2\nabla_{\bot}^{2}A_+ +2\partial_{+}\Phi=j^-~,
\label{eq:E3.6}\end{eqnarray}  
\begin{eqnarray}     
-4\partial_{-}^{2}A_+ + 2\partial_{-}\Phi=j^+~.
\label{eq:E3.7}  
\end{eqnarray}  
The last of these equations has no ``time'' derivative $\partial_{+}$
and is a constraint equation which expresses Gauss' law with the
charge density $j^+$. One can easily transform the system of dynamical
equations (\ref{eq:E3.4})-(\ref{eq:E3.6}) into three independent
equations, Eq.~ (\ref{eq:E3.4}) and 
\begin{eqnarray}   
(4\partial_{+}\partial_{-} -\nabla_{\bot}^{2})(\partial_{-}\Phi) =
(\partial_{+}\partial_{-} -{1\over 2}\nabla_{\bot}^{2})j^\phi
+\partial_{-}j^-= - \partial_{+}\partial_{-}j^\phi-
{1\over 2}\nabla_{\bot}^{2}\partial_{+}j^+ ~,
 \label{eq:E3.8} \end{eqnarray} 
and
\begin{eqnarray} 
(4\partial_{+}\partial_{-} -\nabla_{\bot}^{2})(\partial_{-}A_+)
= {1\over 2}j^\phi + \partial_{-}j^-= - {1\over 2} \partial_{+}j^+ 
+{1\over 2}\partial_{-}j^- ~.
\label{eq:E3.9}\end{eqnarray} 
Here, the right hand side is given in two forms, the original one, and
after its transformation, the one that accounts for charge conservation, 
\begin{eqnarray}   
j^\phi+ \partial_{-}j^- + \partial_{+} j^+=0~ .
\label{eq:E3.10}\end{eqnarray} 
The second form is very useful since it helps clarify
the structure of the field created by the external source. 

For the static source with $\partial_{+} j^+(x)=0$ 
(or $j^+(k)\sim \delta(k^-)$) the derivative $\partial_{-}$ can be easily
removed from  both sides of the Eqs.~(\ref{eq:E3.8}) and (\ref{eq:E3.9})
and thus, no pole $(k^+)^{-1}$ can appear. In this case, the 
equations of motion lead to the diagonal retarded propagator,
\begin{eqnarray}   
 A_{i}(k)=-{j_i(k)\over k^+(k^--i\epsilon)- k_{\bot}^{2}}~,~~i=1,2,+~.
\label{eq:E3.11}\end{eqnarray}
while the static field has to be recovered
via  Gauss' law (\ref{eq:E3.7}) and has only two transverse components:
\begin{eqnarray}   
 A_{r}^{(stat)}(k)={k_r\over k^+ k_{\bot}^{2}}j^+(k^+,{\vec k}_{\bot})~,
\label{eq:E3.12}\end{eqnarray}   
and without any prescription for the pole $(k^+)^{-1}$. The source,
independent on $x^+$, can depend on $x^0$ and $x^3$ in a single combination,
$x^-=x^0-x^3$, and therefore, should propagate {\em without longitudinal
dispersion} at the speed of light in the $x^+$-direction. 
Expression (\ref{eq:E3.12}) is nothing but the Williams-Weiszacker field
of this source.
If $\partial_{+} j^+(x)\neq 0$ then the system of equations
(\ref{eq:E3.4}), (\ref{eq:E3.8}) and (\ref{eq:E3.9}) can be explicitly
integrated to
\begin{eqnarray}   
 A_{i}(k)={1\over k^+(k^--i\epsilon)- k_{\bot}^{2}}
\bigg(-j_i(k)+{k_i\over k^+}j^+(k)\bigg)~,~~i=1,2,+~.
\label{eq:E3.13}\end{eqnarray}  
and, as before, the pole $(k^+)^{-1}$ is due to the integration $dx^-$
that recovers the vector potential via the electric field. No prescription 
can be justified for this integration.  If we assume  that the static
source is propagating at the speed of light and
take $j^+(k)\sim \delta(k^-)$, then we  recover the
Williams-Weiszacker formula  (\ref{eq:E3.12}) as the limit of the full
propagator.

The pole $(k^+)^{-1}$ encounters various types of
physical singularities. One of them is the source with no $x^+$ dependence
propagating in the $x^+$-direction.  The second one emerges as the field
pattern corresponding to the residue in the pole $(k^+)^{-1}$, providing
it is accessible in the calculations. In configuration space,
this pattern is independent of $x^-$ and therefore, propagates at the
speed of light without longitudinal dispersion in the $x^-$-direction.
This field, since it is off-mass-shell, corresponds to the smallest Feynman
$x$ and consequently, to the negative rapidities in configuration space.
One more option is that this is a proper field of the back-scattered
ultrarelativistic charge.   In any of these cases, we are dealing
with  the bounded systems of charge and its static field, propagating in
the null-plane direction which should be renormalized to their physical
parameters.

This simple observation explains the source of controversy which was found
in Refs.~\cite{Raju1,Raju2}; evaluation of the gluon propagator in the gauge
$A^+=0$ has led to the propagator of the gauge $A^-=0$. In fact, these
two gauges are complementary. As long as the theory is supposed to describe
the process of measurement, it unavoidably deals with the physical
singularities on two null-planes.  This is a clear manifestation of the
strong localization of the entire process at its initial moment which
points  to the wedge form of dynamics as a picture which incorporates this
most important physical feature of any measurement at extremely high
energies.

 \section{Physical discussions of the spurious poles}\label{sec:SN4}   

It was already mentioned in the Introduction that the spurious poles are
mostly safe in calculations of the observables like cross sections. In
exceptional cases when they are mathematically
dangerous, the gauge poles of the
propagators were shown to have a clear  physical  meaning; the gauge field
approaches a static configuration. The gauge poles are entirely due to
the longitudinal constituent of the gauge field which is not a dynamical
variable. Therefore, the real problem is to trace the physical origin of 
the static configuration and identify the physical object it belongs to.
Then the natural remedy may be {\em renormalization}, {\em i.e.},
 the brute force
identification of the dangerous element with the physical object of
known properties.

Let us begin with the two popular examples from  QED when this kind of
strategy has proved to be fruitful.  If an electron emits a long wave
photon, then in the limit of $\omega \rightarrow 0$, the photon is 
inseparable from the proper field of the electron. In this case, the formal
divergent perturbation series is assembled (renormalized)  to form the
classical field of the electron \cite{Glauber}.  The reaction of radiation
becomes  negligible and the electric current $j$ should be treated as a
$c$-number rather than an operator. 

When the electronic or muonic pair is created with low relative momenta
the whole series which describes the soft emission should be summed to
form the Coulomb field of the charged pair. In this case, the study ends
up by replacing the plane waves of the final state with the states of
scattering in the Coulomb field \cite{Sakharov}.

More examples may be found in QCD calculations of the deep inelastic e--p
scattering or hadronic collisions. In these cases, it is common to use the
infinite momentum frame where the proton has only $P^+$ component of the
momentum and to connect the gauge condition with the axis $x^+$: $A^+=0$. 
The poles like $1/k^+$ of the null-plane polarization sum eventually enter
the splitting kernels of the DGLAP equations \cite{DGLAP}. They are due to
different processes and are treated differently. The poles $(1-z)^{-1}$~~
$(z=p^+/k^+)$
which appear in the kernels $P_{qq}$ and $P_{gg}$ come from the final
state gluons. They are  treated according to the so-called
(+)-prescription.  
\begin{eqnarray}  {f(z)\over 1-z} \rightarrow
\bigg({f(z)\over 1-z}\bigg)_+  + c\delta (1-z) = {f(z)-f(1) \over 1-z} +
c\delta (1-z) ~~. \nonumber\end{eqnarray}    

The principal value treatment of the pole shields the collinear
singularity, telling us that the emitted forward  gluon is a part of the
proper field of the proton.  The $\delta$-counterterm indicates that this
``emission'' does not change  the quantum numbers of the  proton. Overall,
this procedure is really a kind of {\em renormalization} of the proton wave
function in the environment of the strongly localized interaction with the
electron.
The second type of pole, $1/z$, is treated in different way. This
pole appears in the kernels $P_{qg}$ and $P_{gg}$ and is due to the
retarded tree propagators between consequtive emissions. 
It also reflects the dynamics of the longitudinal gluon field.
Unless we discuss the problem of unitarity, there is no need to screen
it. Including the fusion process into the equations of the QCD evolution
naturally leads to the saturation of the evolution rate. 
 The QFK approach \cite{QFK} describes the QCD evolution as a (virtual)
sequential in real temporal scale process \cite{QGD}; every
act of emission has the preceeding and the subsequent configurations of
the longitudinal field as the boundary conditions. Therefore, a proper
treatment of the longitudinal fields dynamics at the intermediate stage
of the deeply inealastic process is imperative. In fact, this dynamics
is responsible for the low-$x$ enhancement of the structure functions.
 
\section{Conclusion}\label{sec:SN5}

One may view the spurious poles of the gauges $A^0=0$ and $A^+=0$ as an 
artifact of the global choice of the gauge.  In the class of  pure
scattering problems this point of view seems to be, though narrow, but quite
appropriate. If the intermediate dynamics of the system is a subject of
physical analysis,  the choice of the gauge condition still cannot affect
the observables, but it explicitly affects the definition  of the physical
states of the gauge field. 
In this case, a specific gauge may be profitable as long
as it can help single out the physically important details. The
temporal axial gauge explicitly reveals the static  field
configurations by its spurious poles.  The poles of the null-plane gauge
correspond to the proper fields of the ultrarelativistic charged
particles.

\bigskip  

\noindent {\bf ACKNOWLEDGEMENTS}

\bigskip  

I am grateful to D. Dyakonov, B. Muller, L. McLerran, E.Shuryak, and
R.Venugopalan  for many stimulating discussions.    
I much profited from intensive discussions during the International
Workshop on Quantum Chromodynamics and Ultrarelativistic Heavy Ion
Collisions at the ECT${^*}$ in Trento, Italy.

This work was supported by the U.S. Department of Energy under Contract 
No. DE--FG02--94ER40831.

\end{document}